# A RADIO FREQUENCY NON-RECIPROCAL NETWORK BASED ON SWITCHED LOW-LOSS ACOUSTIC DELAY LINES

Ruochen Lu, Tomas Manzaneque, Yansong Yang, Anming Gao, Liuqing Gao and Songbin Gong
University of Illinois at Urbana-Champaign, USA

This work demonstrates the first non-reciprocal network based on switched low-loss acoustic delay lines. A 21 dB non-reciprocal contrast between insertion loss (IL=6.7 dB) and isolation (28.3 dB) has been achieved over a fractional bandwidth of 8.8% at a center frequency 155MHz, using a record low switching frequency of 877.22 kHz. The 4-port circulator is built upon a newly reported framework by the authors [1], but using two in-house fabricated low-loss, wide-band lithium niobate ($LiNbO_3$) delay lines with single-phase unidirectional transducers (SPUDT) and commercial available switches. Such a system can potentially lead to future wide-band, low-loss chip-scale nonreciprocal RF systems with unprecedented programmability [1].

Microwave frequency non-reciprocal networks, e.g. circulators and isolators, have been investigated for full-duplexing radios [2]. Non-reciprocity is conventionally achieved by Faraday effect in ferrite materials [3]. Recently, magnet-free non-reciprocal systems based on modulation of reactance or conductance have been demonstrated [4]–[7]. Despite promising performance, these demonstrations require either a physically large structure for long delays or a high-frequency modulation signals due to the fast phase velocity of electromagnetic waves. Moreover, the bandwidth of non-reciprocity is limited by the modulation frequency and required phase matching condition. To overcome their limitations, we harness shear horizontal acoustic waves in a $LiNbO_3$ thin film to produce long delays (280 ns) with sub-4 dB IL over 1-mm size. Combining with our frequency-independent framework, this work has achieved wideband non-reciprocity employing unprecedentedly low temporal effort (e.g. frequency and depth).

The schematic of the 4-port non-reciprocal system [Fig. 1(a)] consists of two delay lines and four single pole single throw switches. The switches are controlled by four control signals [Fig. 1(b)], with a period (4δ) that is four times the delay line's group delay. Control signals on opposite sides of the delay lines are offset by δ. In operation, the signals flowing into Port 1 are time-multiplexed onto the two delay lines and subsequently de-multiplexed to Port 2 by turning on the switches connected to Port 2 δ time after the signals launched from Port 1. The time-reversal symmetry is broken through sequentially timing the switching from one side of the delay lines to the other side. Consequently, signals fed to Port 2 are rejected by Port 1's closed switches and received by Port 3. The assembled circulator performance is simulated [Fig. 5(b)] with the control signal frequency set to 877.2 kHz to match the group delay. An IL of 5.6 dB and an isolation of 30 dB is obtained.

Experimentally, we implemented two standalone switch boards and one delay line board, and assembled them as the circulator seen in Fig. 1(c). The switch board design schematic and the constructed board are shown in Fig. 2. On the delay line board, a pair of in-house fabricated SPUDT [8] $LiNbO_3$ acoustic delay lines [Fig. 3(a)-(c)], were wirebonded to LC matching networks [Fig. 3(d)-(e)]. Measured and simulated S-parameters and group delays of the delay line board are shown in Fig. 4. 4 dB IL and around 280 ns group delay are measured. As shown in Fig. 6, the measured S-parameters exhibit great performance symmetry between ports, a minimum insertion loss around 6.7 dB, and an isolation larger than 27 dB over a bandwidth of 13.6 MHz (8.7% FBW). Currently, the loss is limited by insufficient directionality in the SPUDT design and impedance matching network, which will be significantly reduced by further optimization on acoustic delay lines. The spectral contents of different ports are measured when port 1 is excited by a single tone (Fig. 7). The intermodulated tones are caused by the non-ideal switching and multi-reflection on the delay lines exist in the spectrum, which can be significantly diminished using a differential structure [9] in future work.

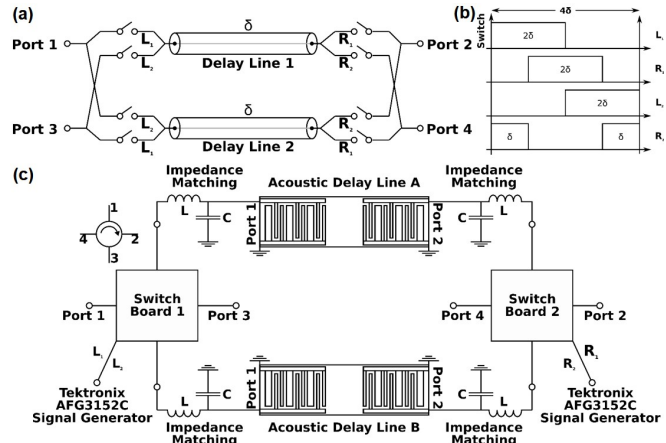

*Figure 1: (a) Schematic of 4-port circulator based on switched delay lines. (b) Switch control waveforms applied to the network for producing nonreciprocal response. (c) Block diagram of the constructed 4-port circulator, including switching modules, impedance matching networks, and unidirectional acoustic delay lines.*

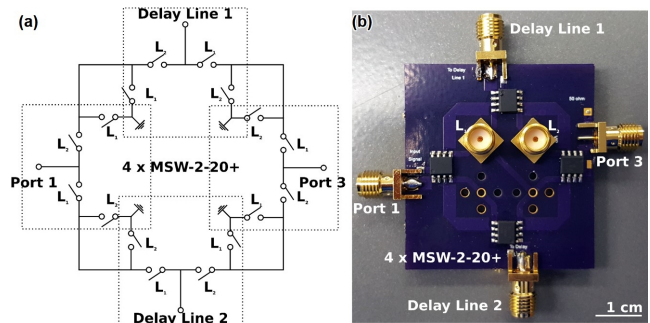

*Figure 2: (a) Schematic of the switching module with labeled components and interfaces. (b) Implemented switching module.*

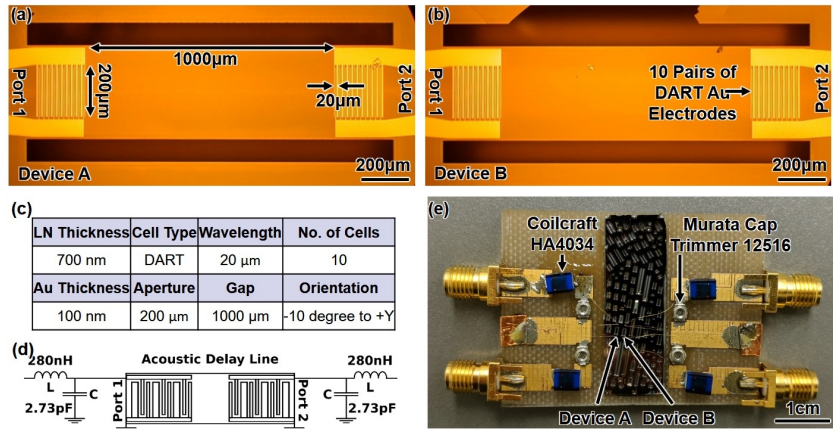

*Figure 3: (a)-(b) Fabricated acoustic delay lines with single-phase unidirectional transducers (SPUDT). A pair of distributed acoustic reflection transducers (DART) are arranged on both end of the suspended LiNbO$_3$ thin film. Design parameters are listed in (c). For lowering the insertion loss of the delay lines, inductor-capacitor (LC) circuits are used to match the impedance to 50 Ω, as seen in (d). (e) delay lines assembled with matching networks on a FR-4 board.*

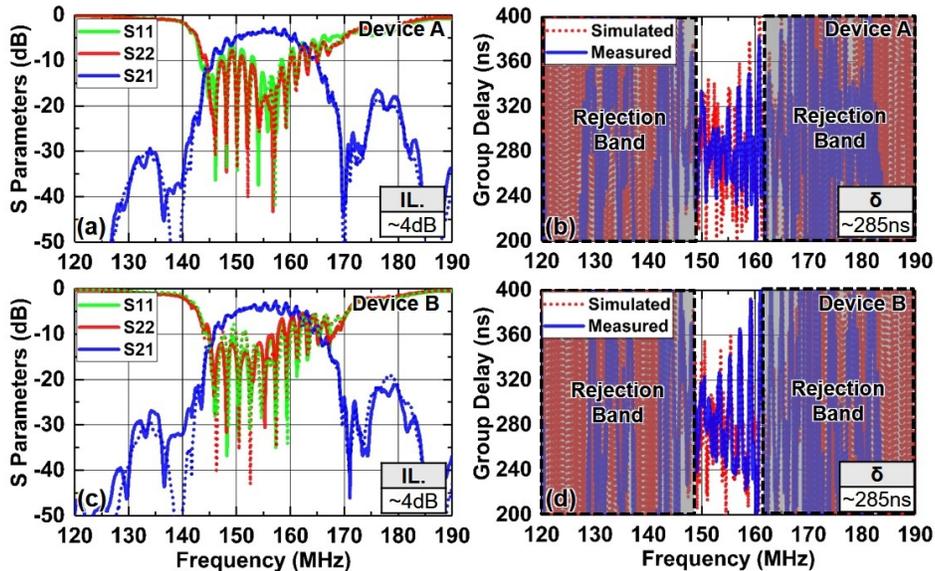

*Figure 4: Measured (solid) and simulated (dashed) S-parameters and group delay of the matched (a)-(b) delay line A and (c)-(d) delay line B. An insertion loss around 4 dB and a group delay of 280 ns are measured. In the simulation, acoustic delay lines are represented by their measured S-parameter performance.*

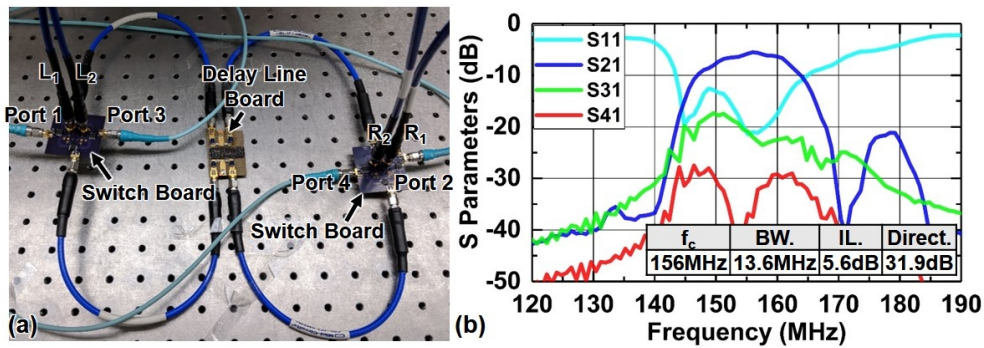

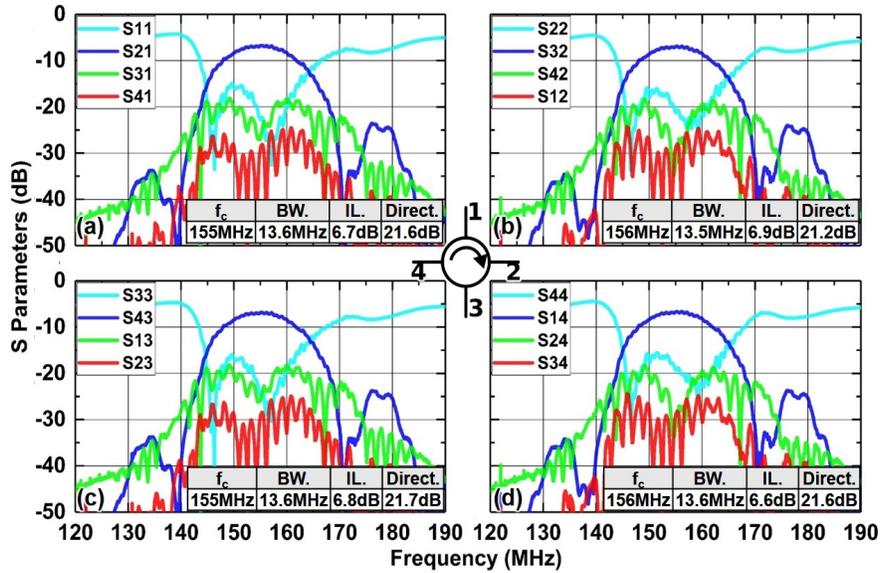

*Figure 5: (a) Experiment setup of 4-port circulator, consisting of 2 switch boards and 1 delay line board. Interfaces are labeled on the figure. (b) Simulated S-parameters obtained from Advanced Design System. In the simulation, 2ns switching time and on-state insertion loss of the switches are considered. **Control signals are set to be 877.2 kHz (1.14 μs period).***

*Figure 6: Measured S-parameter performance of the 4-port circulator. Great performance symmetry is shown in the measurement. Minimum insertion loss (IL) around 6.7 dB is measured at different ports. A bandwidth of 13.6 MHz (8.7% fractional bandwidth) is obtained. Directivity larger than 21 dB is obtained between the forward and backward propagation path (e.g. between $S_{12}$ and $S_{21}$). Return loss is better than 15 dB at each port. The measurement is carried out at -10 dBm power level. Control signals are set to be 877.2 kHz (1.14 μs period).*

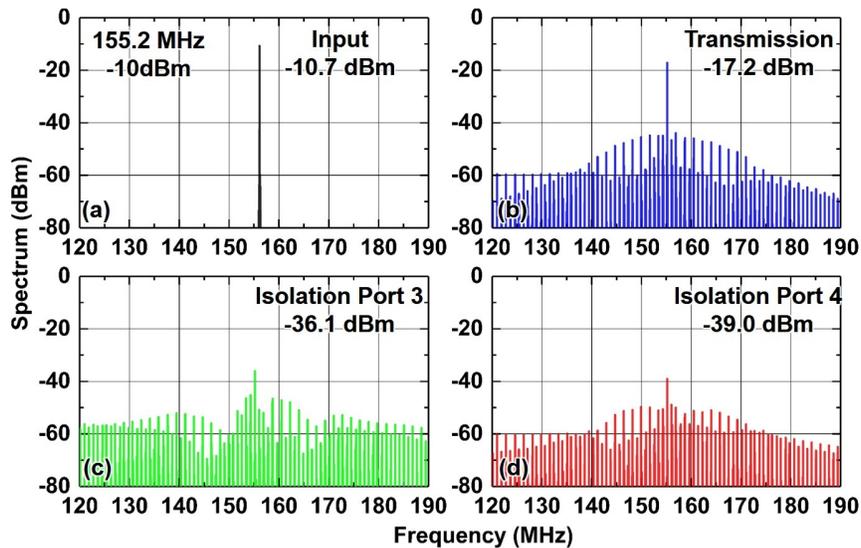

*Figure 7: Measured spectral content of (a) input signal, (b) transmitted signal at port 2, showing 6.5 dB insertion loss, (c) transmitted signal at port 3, indicating 25.4 dB isolation at port 3, and (d) transmitted signal at port 4, indicating 28.3 dB isolation at port 4. The intermodulation is caused by the non-ideal switching and multi-reflections in the spectrum.*